\documentclass[12pt]{iopart}
\usepackage{graphicx}

\usepackage[active]{srcltx}
\usepackage{subfigure}
\usepackage{pifont}
\usepackage{cite}

\newcommand{\remark}[1]{}

\newcommand{\Ca}{$^{40}$Ca$^+\,$}

\usepackage{iopams}
\begin{document}

\title[Electric field compensation and sensing with a single ion in a planar
trap]{Electric field compensation and sensing with a single ion in a planar
trap}

\author{S.~Narayanan$^{1,2}$, N.~Daniilidis$^{1,2}$, S. A. M\"oller$^{1,2,3}$,
R.~Clark$^{2,4}$, F. Ziesel$^{5}$, K. Singer$^{5}$, F.
Schmidt-Kaler$^{5}$, and H.~H\"affner$^{1,3}$}

\address{$^{1}$ Dept. of Physics, University of California, Berkeley, CA 94720,
USA\\}
\address{$^{2}$ Institut f\"ur Quantenoptik and Quanteninformation, Innsbruck,
Austria\\}
\address{$^{3}$ Materials Sciences Division, Lawrence Berkeley National
Laboratory, Berkeley, CA 94720, USA\\}
\address{$^{4}$ Center for Ultracold Atoms, Massachusetts Institute of
Technology, Cambridge, MA, USA\\}
\address{$^{5}$ Institut f\"ur Physik, Universit\"at Mainz, Mainz, Germany\\}

\ead{hhaeffner@berkeley.edu}

\begin{abstract}
We use a single ion as an movable electric field sensor with accuracies on the
order of a few V/m. For this, we compensate undesired static
electric fields in a planar RF trap and characterize the static fields over an
extended region along the trap axis. We observe
a strong  buildup of stray charges around the loading region on the trap
resulting in an
electric field of up to 1.3~kV/m at the ion position. We also find that the
profile of the stray
field remains constant over a time span of a few months.
\end{abstract}

\section*{Introduction}

Laser cooled trapped ions offer a very high level of control, both of their
motional and internal quantum states.
At the same time, the large charge-to-mass ratio of ions makes their motion very
sensitive to electric fields, both static
and oscillatory. Thus, trapped ions recently emerged as a tool in small-force
sensing\cite{Maiwald2009a, Biercuk2010}. 
More common applications of trapped ions are in quantum information
science \cite{Wineland1998,Home2009} and frequency metrology
\cite{Schmidt2005,Rosenband2008}. All these applications
can benefit from scalable ion-trap architectures based on microfabricated ion
traps. 

In particular, a promising route to achieve scalable quantum information
processing uses
complex electrode structures\cite{Wineland1998,Kielpinski2002}. Considerable
effort is made in
developing microfabricated trap architectures on which all trap electrodes lie
within one plane
\cite{Chiaverini2005,Seidelin2006,Britton2006,Pearson2006,Labaziewicz2008,
Leibrandt2009,Allcock2010,Amini2010}. These so-called
planar traps facilitate creation of complex electrode structures and are, in
principle, scalable to large numbers of electrodes. Moreover, this approach makes
use of mature microfabrication technologies and is ideally suited to approaches
involving hybrid ion-trap or solid state systems \cite{Tian2004, Daniilidis2009}.

Despite the advantages of planar trap architectures, a number of issues remain unsolved. 
To achieve reasonably large trap frequencies, planar traps
require shorter ion-electrode distances than conventional three-dimensional
traps
\cite{Chiaverini2005}. This results in high motional heating rates for the ions
\cite{Turchette2000,DesLauriers2006a,Daniilidis2011} and
causes charge buildup via stray light hitting the 
trap electrodes\cite{Harlander2010}. In addition, the proximity of the charges 
increases the effect of charge buildup as compared to macroscopic three
dimensional traps.
Finally, planar traps do not shield stray electrostatic fields from the
environment surrounding
the trap as well as the three dimensional trap geometries tend to do.
Combined, these effects make the operation of planar traps much more sensitive to uncontrolled charging effects.

To harness the full advantages of segmented ion traps, ion-string splitting and
ion shuttling operations
are required\cite{Home2009,Blakestad2009}. For the reliable performance of these
operations, control of the electrostatic 
environment over the full trapping region is necessary. Typically one employs
numerical electrostatic solvers to determine the potential
experienced by the ions and generates electrode voltage
sequences that will perform the desired ion shuttling
\cite{Schulz2006,Schulz2008}.  Stray electrostatic fields, however, displace
the ions from 
the RF-null of the trap and thus introduce so-called
micromotion\cite{Berkeland1998} sometimes to the point where trapping 
is no longer feasible. 
Thus, precise characterization and compensation of stray electric fields in the
trapping region is required.
 
Conventional methods to sense and compensate the electric stray fields cannot
easily be extended to
planar traps because typically the stray fields are quantified via the
the Doppler shift induced by the micromotion.
It is undesirable to scatter UV light from the trap electrodes, and, thus,
for planar traps, the
detection laser typically does not have a sizable projection on the motion
perpendicular to the plane of the trap. 
We address these issues by applying a new method to compensate for stray fields
well suited for planar trap geometries \cite{Rosenband2009,Ibaraki2011}.

Based on the voltages required to compensate the stray fields, we realize a
single-ion electric field
sensor characterizing
the electric stray fields along the trap axis.  
We observe a strong buildup of stray charges around the loading
region on the trap. We also find that the profile of the stray field remains
constant over a time span of a few months. The strength of the electric stray
fields and its position on the trap is 
correlated with the high heating rates observed close to the loading
region \cite{Daniilidis2011}.
 
\section*{Experimental Setup}
\begin{figure}[trap]
\begin{center}
\includegraphics[width=0.4\textwidth]{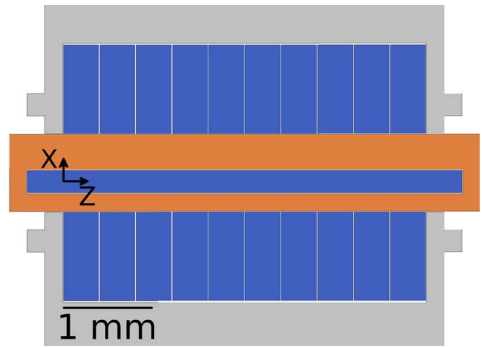}
\end{center}
\caption{\label{fig:trap}
Schematic of the trap used for the measurements\cite{Daniilidis2011}. The
DC electrodes are drawn in blue, the RF electrode in orange, and the ground
plane in gray. Details of the bonding pads to the DC electrodes are not shown
for simplicity. The axes indicate the origin of the coordinate system. The green
line along the z axis on the central DC electrode indicates the range of axial
positions in which the stray electric fields shown in Fig.\,\ref{fig:el-field}
were measured. The circular mark on this line indicates the location used as a
loading region, around which the highest increase in stray electric fields was
observed.}
\end{figure}

We use a planar trap with gold electrodes deposited on a
sapphire substrate to trap single \Ca ions at a height of 240~$\mu$m above the
trap plane, see Fig.\,\ref{fig:trap}. Ions are created via two step photoionization from a neutral calcium
beam using 250~mW/cm$^{2}$ of laser light at 422~nm  and 750~mW/cm$^{2}$
of laser light at 375~nm. Both the laser beams are focused to a waist size of 50~$\mu$m. Great care has been taken to minimize exposure of the
trap surface to the neutral calcium beam.

The RF electrode is driven at a frequency $\Omega/2\pi \approx$ 15 MHz,
amplified to $100\,$mW and stepped up via a helical resonator in a quarter wave
configuration to a voltage of approximately 100 V amplitude. A 2:1 asymmetry in the width
of the RF electrode results in a tilt of the radio frequency quadrupole by
approximately $16\,^{\circ}$ in the $XY$ plane. The DC electrodes are used to
move
the ion along the axial direction and to compensate the stray fields. The DC
voltages
used for trapping and compensation are between -10 V and 15 V. Typical secular
frequencies in this work were $(f_{x'}, f_{y'}, f_{z}) \approx $ (1.2, 1.4, 0.4)
MHz where the primes refer to the frame of reference rotated by $16\,^{\circ}$.
 
For Doppler cooling and detecting the ions, we use a diode laser at 794\,nm,
which is frequency doubled using a ring cavity to produce a wavelength of
397\,nm. A second diode laser at 866\,nm is used as a repump. Both lasers are
frequency locked to cavities using the Pound-Drever-Hall method, and their
frequencies can be varied by changing the cavity lengths with piezoelectric
elements. The intensity of the detection laser at 397\,nm is adjusted to about
40\,${\rm mW / cm^2}$ and the intensity of the repump laser at 866nm is adjusted
to approximately 120\,${\rm mW / cm^2}$. The Doppler cooling and repump lasers
are overlapped and sent to the trap using a photonic crystal fiber. The laser
beam is aligned almost parallel to the surface of the trap and forms an angle
of approximately $45\,^{\circ}$ with the $X$ and $Z$ axes. Ion fluorescence is
collected perpendicular to the trap plane using a lens system of NA = 0.29 and
split between a PMT and CCD
camera on a 90:10 beam splitter.
 
\section*{Minimizing micromotion}
\label{sec:micromotion-compensation}
In an ideal Paul trap the ion is confined to a position at which the electric field due to the oscillating drive voltage on the RF electrodes is zero.
Stray DC electric fields, however, push the ion off the RF node and the ion undergoes
so-called micromotion driven by the oscillating RF field \cite{Berkeland1998}.
This motion causes broadening of the electronic transitions of the ion and,
among
other things, leads to a higher temperature limit for Doppler cooling
\cite{Wineland1998}. In addition, micromotion can lead to the heating of trapped ions due to the noise present at the secular sidebands of the micromotion drive \cite{Wineland1998a, Blakestad2009}. In order to position the ion on the
RF node, the DC potential is carefully adjusted to minimize micromotion. Crucial
to all minimization schemes is the efficient detection of micromotion in all
three spatial directions.

Different techniques exist for fine-tuning the compensation of micromotion. 
The photon correlation method\cite{Berkeland1998,Allcock2010} relies on
correlating the
ion fluorescence to the phase of the RF field. In the resolved sideband
method\cite{Berkeland1998, Schulz2008}, the sidebands of a narrow atomic
transition are compared with the carrier transition to estimate the modulation
index. Both
methods are widely used to suppress micromotion in 3D traps. However, neither
method can be easily
extended to surface Paul traps
because they directly use the Doppler shift induced by the ion motion. In
the case of surface traps, the geometry typically limits the laser alignment to be in the
plane parallel to the trap surface resulting in no Doppler shift associated with
the oscillations perpendicular to the trap surface unless one directs laser
light onto
the trap electrodes. However, it has been documented that UV light hitting the
trap surface can lead to dramatic charge buildup even to the point where the
trap becomes inoperable for days\cite{Harlander2010}. To circumvent this
obstacle, the infrared repump light in \Ca has been used to detect the Doppler
shift
perpendicular to the trapping plane for micromotion
compensation \cite{Allcock2010}. However, many ion species such as Mg$^+$,
Al$^+$, Hg$^+$, Cd$^+$, Be$^+$ do not have such transitions in the infrared and
other methods need to be employed.

Here we compensate the micromotion perpendicular to the trap plane with the
following method.
When the ion is displaced from the RF node, any voltage applied on the RF
electrode creates an electric field at the ion position. If this voltage contains a frequency component which is in resonance with one of the ion secular frequencies the ion can get excited
in the direction of the secular mode provided that the driving field from the RF
electrodes has some projection \cite{Blakestad2009}.
Experimentally we find that large oscillation amplitudes of each of the three
secular motions can be detected as a drop in ion fluorescence. 
The dynamics of ion fluorescence in the
presence of the cooling laser and a resonant excitation are complex and go
beyond the scope of this study \cite{Akerman2010}.
Minimizing micromotion is achieved by shifting the ion position via DC
potentials until the ion is in the
RF minimum and cannot be excited by driving the RF electrode at any of the
secular frequencies. This method is also being used by NIST \cite{Rosenband2009} and Osaska \cite{Ibaraki2011} ion trap groups.

\subsection*{Implementation}
In order to implement this method, we first position the ion within
1~$\mu$m along the $X$ direction 
from the RF null using the CCD camera and varying the RF amplitude. Further
compensation in the $X$ direction 
is achieved by reducing the linewidth of the S$_{1/2}$-P$_{1/2}$ transition. For
this, the detection laser intensity is adjusted close to saturation and red-detuned from the transition so
that the fluorescence drops to half that of the value at the resonance. Then 
compensation voltages are adjusted to minimize the fluorescence.
Both methods detect only micromotion along the {\it
X } direction, i.e. the direction which is parallel to the trap surface.
For a very coarse compensation along the {\it Y }
direction (perpendicular to the trap surface), we keep the
frequency of 397\,nm and 866\,nm laser on resonance and maximize the ion
fluorescence by adjusting the compensation voltages.

\begin{figure}[micromotion]
\centering
\subfigure[]{\label{fig:MMdip}\includegraphics[width=0.2\textwidth]{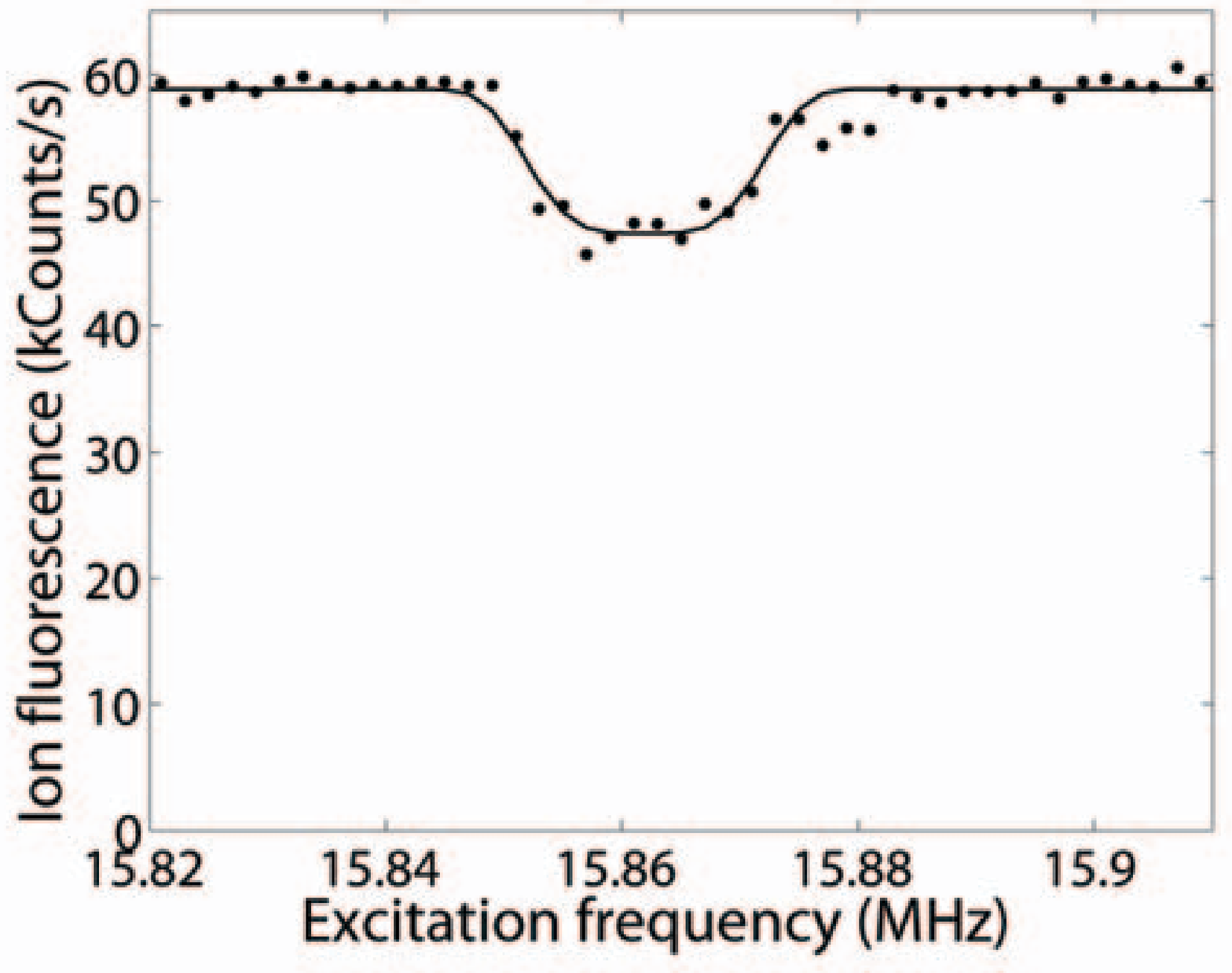}}\\
\subfigure[]{\label{fig:MMx2d}\includegraphics[width=0.2\textwidth]{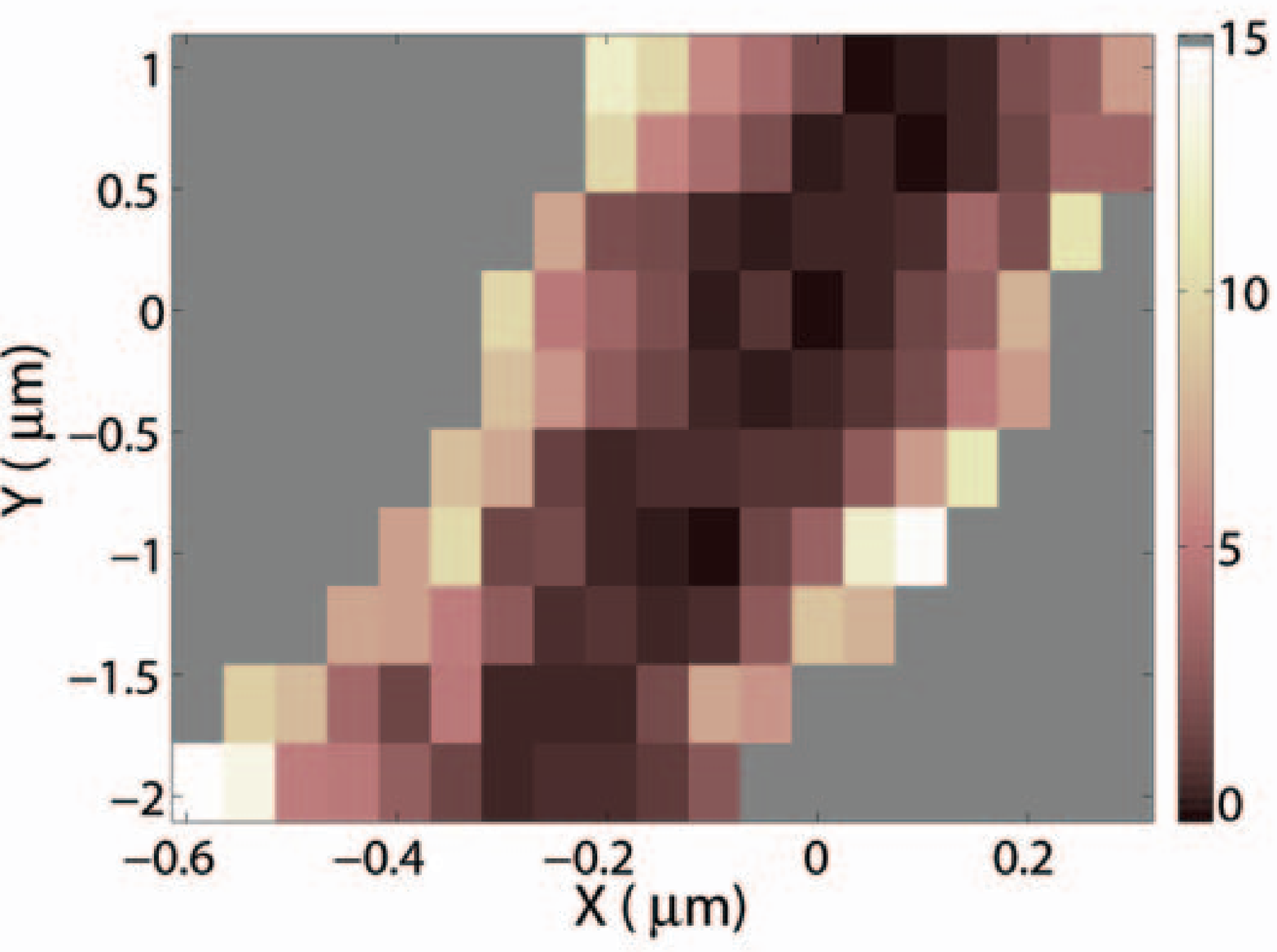}}
\subfigure[]{\label{fig:MMy2d}\includegraphics[width=0.2\textwidth]{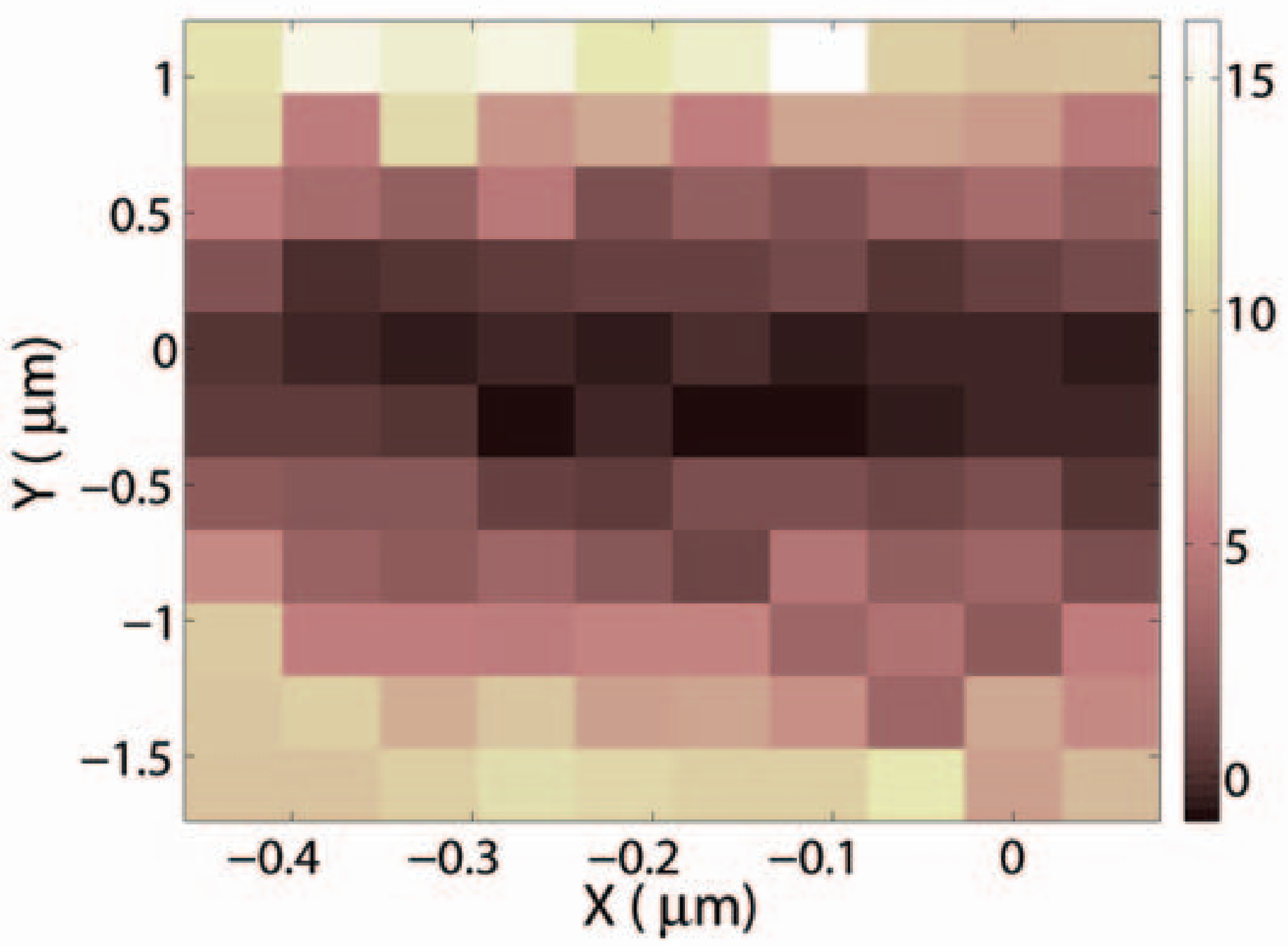}}
\subfigure[]{\label{fig:MMx1d}\includegraphics[width=0.2\textwidth]{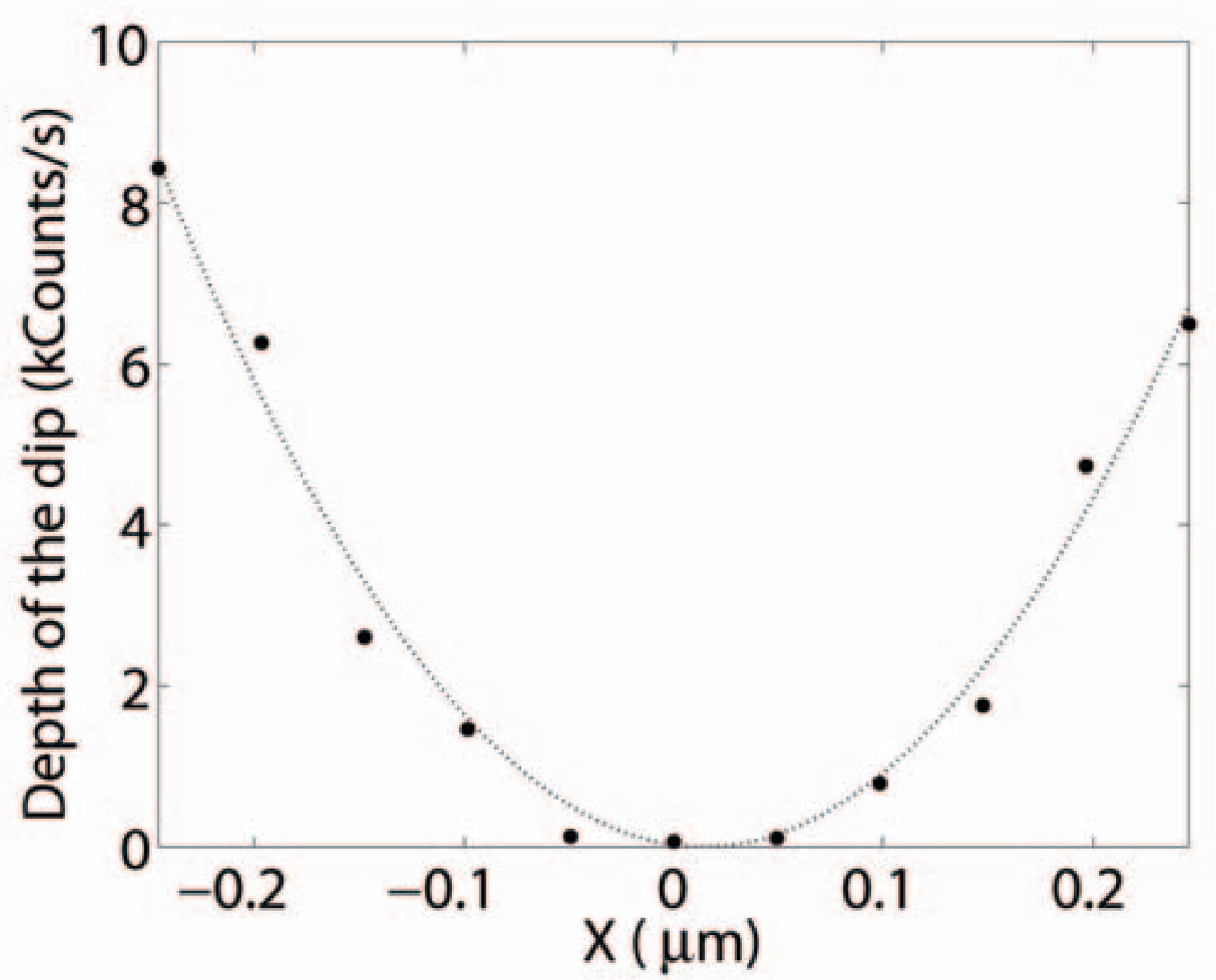}}
\subfigure[]{\label{fig:MMy1d}\includegraphics[width=0.2\textwidth]{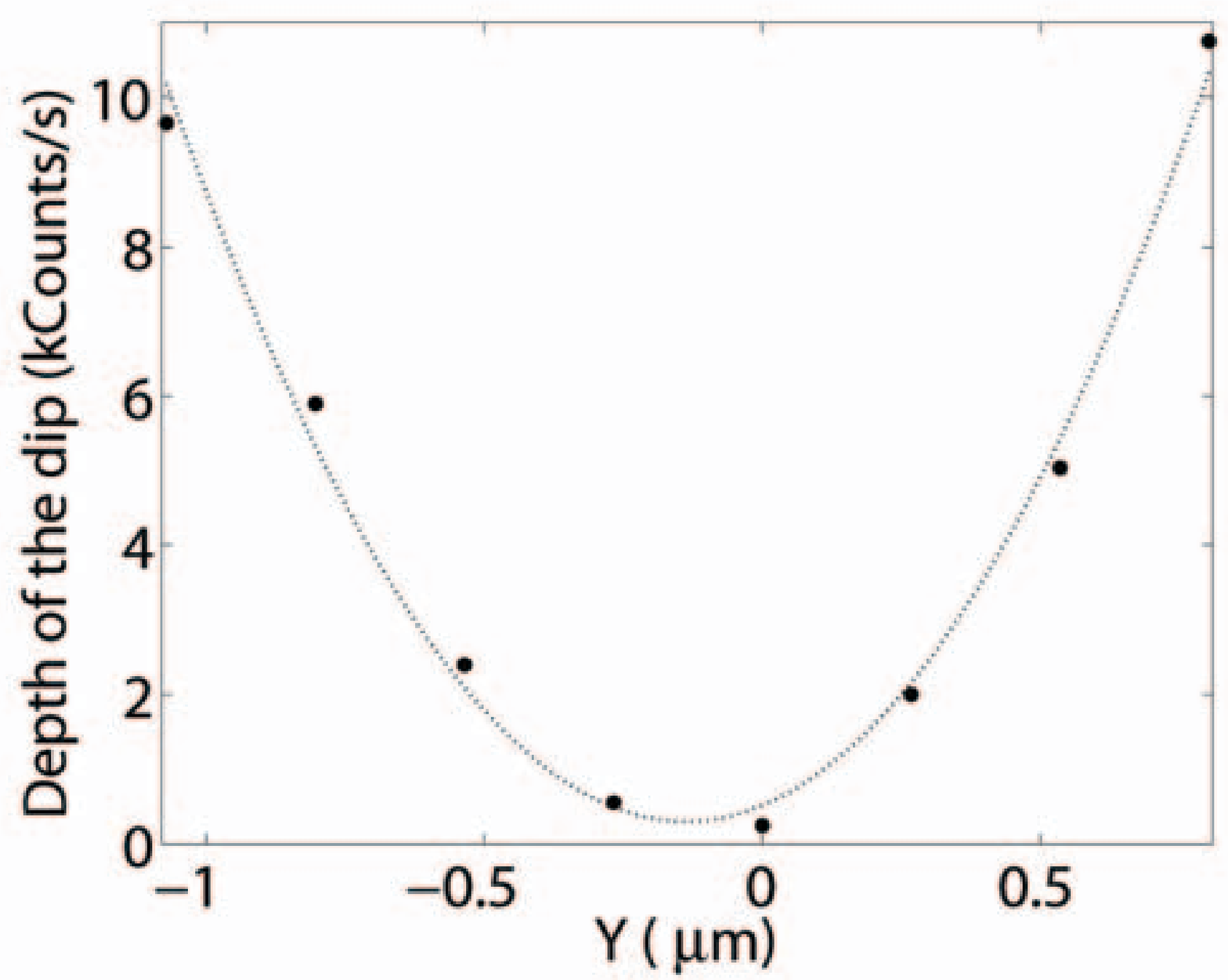}}

\caption{\label{fig:MMcomp}
Micromotion compensation measurements. (a) Drop in ion
fluorescence when the frequency of the additional excitation is equal to one of
the secular frequencies of the ion, the data is fitted using
$A_0(1-D)\left(e^{-[\frac{(x-x_0)^2}{a^2}+\frac{(x-x_0)^4}{b^4}]}\right)$ for a good evaluation of
the dip-depth $D$,  (b)\&(c) Change in the dip depth for the $X$ and $Y$ sideband, respectively,
as a function of the ion position,
(d)\&(e) a cross section of (b)\&(c) respectively fitted with a parabola. Gray
area in (b) is region without data since the excitation might drive the ion out
of trap. The darker the color the less excitation, i.e. less dip-depth was
observed.}
\end{figure}

Once a coarse compensation is achieved using the above methods, we proceed with
the method as outlined in the beginning of this section. Instead of exciting the
secular frequency $\omega_{i}$ directly, we excite at a frequency $\Omega +
\omega_{i}$ \cite{Blakestad2009, Wineland1998a}. To achieve this, the excitation signal from a function generator is
mixed with the trap drive $\Omega$ before it is amplified and stepped up with a
helical resonator, and scanned around $\Omega + \omega_{i}$. When the
frequency of the excitation becomes resonant with $\Omega + \omega_{i}$, 
the ion heats up resulting in a decrease in fluorescence
(see Fig.\,\ref{fig:MMdip}), \cite{Ibaraki2011, Blakestad2009}. A crucial requirement is that the
step-up circuit which
produces the high-voltage trap drive has a large enough bandwidth. The bandwidth
of the helical resonator in our experiment is  270\,kHz and allows compensation
with excitation frequencies up to order $\Omega \pm 2\pi\times$2~MHz. The
frequency of the Doppler cooling laser was detuned between 1 and 5~MHz below the
$S_{1/2} - P_{1/2}$ transition to maximize sensitivity of the ion
fluorescence to the ion kinetic energy.

For compensation of micromotion in the {\it X } direction, the excitation signal
is scanned repeatedly around $\Omega + \omega_{x}$ while adjusting the voltages
on the DC electrodes such that between successive scans the static electric
field minimum moves predominantly in the {\it X } direction. The compensated
position is reached when resonance of the excitation does not result in a
decrease of fluorescence. The same process is repeated for compensation in the
{\it Y } direction. Results are shown in Fig.\,\ref{fig:MMcomp}. 
 
Fig.\,\ref{fig:MMx2d} and Fig.\,\ref{fig:MMy2d} show the change in the dip depth
 when the excitation frequency is $\Omega + \omega_{x}$ and $\Omega +
\omega_{y}$, respectively, as a function of ion position along the {\it X} and
the {\it
Y} direction. The gray area in Fig.\,\ref{fig:MMx2d} is the region were no data
is acquired since the excitation would drive the ion out of the trap.
Fig.\,\ref{fig:MMx1d} and Fig.\,\ref{fig:MMy1d} show cross sections of the 2D
plots along the equilibrium positions of the ion when the DC saddle point 
moves along the {\it X} and
{\it Y} directions, respectively. 
The energy gain rate of the ion $\Gamma$ is expected to be $\Gamma\sim E^2$,
where $E$ is 
the strength of the exciting electric field.
We find that the data can be well fitted with a
parabola suggesting
that the dip depth is linear in the ion energy for our experiments while
the electric field $E$ can be described well as a quadrupolar field.

The accuracy of our measurements is estimated by calibrating the ion
displacement as
a function of
change in the DC voltages using the CCD camera in the X direction and the
detection laser in the Y direction. These values were verified by modeling the
displacement of the DC minimum with variation of the compensation voltages.
By translating the applied voltage into actual displacement, we determine an accuracy of about 50\,nm in the X direction and 300\,nm in Y
direction in positioning the ion at the RF minimum. This corresponds to excess
micromotion amplitudes of 6\,nm and 40\,nm in these respective directions. The
accuracies could be further improved by increasing the excitation voltage and
decreasing the frequency detuning of the detection laser from resonance.

One concern with this method is that in practical situations the DC electrodes
pick RF voltage, the phase of which might be shifted, or which might depend on the RF excitation frequency. The reason for this is that the DC
electrodes 
may capacitively couple to the RF electrode, with a frequency dependent coupling
determined by the 
filtering circuits connected to the DC electrodes. In our setup, we estimate the
RF pickup of the 
excitation at $\Omega + \omega_{i}$ on the DC electrodes to differ by less than
20~$\mu$V from the pick
 up at $\Omega$ which would shift the RF-null by about 15~pm, and thus not limit the
accuracy.

\section*{Electric field sensing}
\label{sec:electrostatics}
It is instructive to extract the size and direction of the 
stray fields from the compensation voltages along various trapping positions.
We derive the stray field from the applied compensation voltages via an accurate
electrostatic model of the trapping fields. Thus determining the stray
fields over an extended region yields important information for shuttling
experiments or investigating the mechanism for charge buildup.

We model the trap potentials using the boundary element method solver CPO \cite{CPO}.
Since the radial trap axes are tilted, we determine the radial secular frequencies from 
a 2-dimensional polynomial fit in the
$XY$-plane. The agreement between
experiment and simulation is better than 4\%, corresponding to a disagreement of
less than 20\,kHz for axial secular frequency of 500\,kHz, and 30\,kHz for
radial frequencies of between 1 and 2\,MHz. We attribute the disagreement to
details of the trap electrode geometry which were not included in the
electrostatic simulation, as well as to the unknown spatial inhomogeneity of the
electrostatic stray fields. Only the
spatially inhomogeneous part of the stray field will contribute to the
discrepancy between simulation and experiment, since the electrostatic stray
fields are compensated and included in the simulations, as we now describe.
 
The essential part of the stray field determination is to perform micromotion compensation as
described above. After the compensation parameters have been determined, we measure the ion
position along the Z direction using the CCD imaging system. The
values of the DC voltages at the compensated configuration and the ion position 
are then input to a minimization algorithm. The algorithm finds the stray field 
which results in the observed
compensation parameters and position.
 
The accuracy of our measurement scheme is limited by an estimated uncertainty of $\pm$2.5~$\mu$m in
measuring the absolute ion  
position along the trap axis. This position uncertainty arises from the 
imprecision of the alignment of
the microscope objective used for imaging with respect to the trap plane and
from the size of trap features used to determine the position along the trap
axis. This results in a systematic
error of the electric field curves, mainly an offset of the entire curve. We find
typical inaccuracies of $(\delta E_x,\,\delta E_y, \,\delta
E_z)=\pm(5.5,\,3,\,15)\,$V/m.

The precision in determining the electric field at each ion position is on
the order of a few V/m, and is limited by the precision to which the
compensation parameters and relative ion position are determined. The errors from
imperfect compensation lead to stray field imprecisions of $\pm$0.4~V/m in
the horizontal direction and
$\pm$2.5~V/m in the vertical direction, where we assume that the precision is limited by the ion displacement steps in Fig.\,\ref{fig:MMcomp}.

The accuracy in determining the axial ion position at any given setting with respect
to its positions in neighboring settings is limited by aberrations in the
imaging optics and is estimated to be $\pm$0.1\,$\mu$m. This limitation leads to an
imprecision of $\pm(2,\,4,\,0.5)$\,V/m in the
electric fields. The above limitations of the measurement leads to a total error of $\pm(7,\,10,\,16)$V/m, however, the limitations are of technical
nature and can be further improved.
 
\begin{figure}[!tb]
\begin{center}
\includegraphics[width=0.5\textwidth]{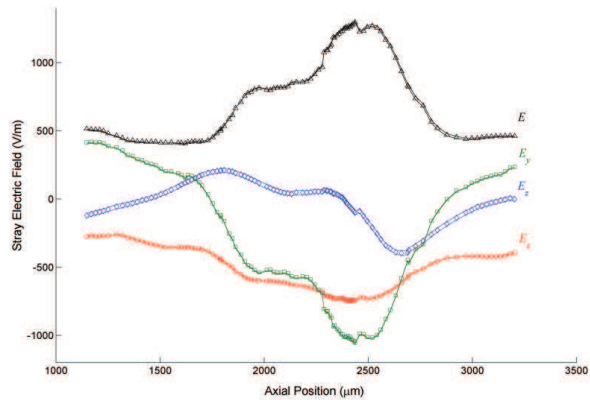}
\end{center}
\caption{\label{fig:el-field}
Stray electrostatic field along the trap axis. The three components $E_x$
($\circ$), $E_y$ ($\square$), $E_z$ ($\Diamond$) and the total electric field $E$
($\bigtriangleup$) are shown. 
The solid curves on the two sides of each curve correspond to the possible
systematic offset arising from the uncertainty in the absolute axial position,
as described in the text.}
\end{figure}
 
\label{sec:electric-stray-fields}
We performed this type of analysis over an extensive part of the trap.
Measurements were performed during several months. The
results obtained in the first weeks of trap installation in vacuum and trap
operation were significantly different than those obtained later on. Initially,
the magnitude of the stray electric fields in the radial directions
varied on a day-to-day basis with a mean of 47 V/m and a  standard deviation of 52 V/m.
Measurements during that time were carried out only for
one isolated axial ion position at 2240~$\mu$m. After two months of trap
operation, there was an abrupt change in stray electric fields to considerably
higher values. This change coincided with a pressure increase in the chamber
 to approximately $10^{-7}$\,mbar because of a temporary ion pump failure. Subsequently, the stray field was
mapped along the trap axis over a length of approximately 2\,mm. The results are shown in
Fig.\,\ref{fig:el-field}. We measured stray fields as large as 1300\,V/m.

Intuitively it is conceivable that two localized charge sources located near the Z axis
are responsible for the observed electric field pattern. 
However, such inverse problems cannot be solved without further assumptions about the charge distribution \cite{Bosco1981, Smith1985}. Attempts to reconstruct the charge distribution mathematically are in progress and will be reported elsewhere 

The position of the postulated charges close to the most frequently used trapping positions 
suggests that exposure of the trap material to laser light 
caused the long term charging. Finally, the high stray fields also appear to be spatially and temporally
correlated with the observed high heating rate\cite{Daniilidis2011}. 
 
Our measurement scheme allowed us to monitor charging of the trap during
operation. We found that the stray fields after the abrupt change in trap
behavior are semi-permanent, showing a slow temporal drift on the order tens V/m
per week. The most striking occurrence of this slow drift is the slight
discontinuity in the data of Fig.\,\ref{fig:el-field} around an axial position of
2450\,$\mu m$. This drift by 70\,V/m, corresponding to a fractional change in the
stray field by $\approx$6\%, occurred over a period of 9~days. Besides the slow
drift of the stray field, short-term charging is observed in the course of a
day. This is typically on the order of 100\,V/m. When the laser light is turned off, discharging occurs over the course of several hours.
 
\section*{Summary and Conclusions}
In this article, we demonstrate a simple, yet
efficient method of measuring stray electric fields in planar ion
traps. This method permits us to sense electric fields over an
extended region, thus providing insight into the undesired charging of
ion traps. This ability to characterize electric fields in the
trapping region will be a valuable tool for evaluating planar ion
traps, for developing ion loading approaches that minimize stray
charging, and for compensating stray fields. We expect that this
technique will be useful not only for scalable quantum
information processing, but also for precision frequency metrology applications of trapped ions.

\section*{Acknowledgment}
The experiments were supported by the Austrian Ministry of Sciences with a START
grant
and by the Director, Office of Science, Office of Basic Energy
Sciences, Materials Sciences and Engineering Division, of the U.S. Department of
Energy under Contract no. DE-AC02-05CH11231. N. Daniilidis was supported by the
European Union with a Marie Curie fellowship. F. Schmidt-Kaler acknowledges 
support from the German-Israel foundation and the EU network AQUTE.

\section*{References}
 
\bibliographystyle{unsrt}

\end{document}